\documentclass[runningheads]{llncs}
\usepackage{orcidlink}
\usepackage[T1]{fontenc}
\usepackage{graphicx}
\usepackage{booktabs}
\usepackage[misc]{ifsym}
\newcommand{\corr}{(\Letter)}
\usepackage{mwe}
\usepackage{amsfonts,amsmath,amssymb}
\usepackage{multirow}

\begin{document}

\title{Which Company Adjustment Matter? Insights from Uplift Modeling on Financial Health}

\titlerunning{Uplift modeling for Company Adjustment Act Analysis}


\author{Xinlin Wang\inst{1} \orcidID{0000-0003-2275-9424} \and
Mats Brorsson\inst{1} \corr \orcidID{0000-0002-9637-2065}}


\institute{Interdisciplinary Centre for Security, Reliability and Trust, University of Luxembourg, L-1855 Kirchberg, Luxembourg\\
\email{\{xinlin.wang,mats.brorsson\}@uni.lu}}
\maketitle              

\begin{abstract}
Uplift modeling has achieved significant success in various fields, particularly in online marketing. It is a method that primarily utilizes machine learning and deep learning to estimate individual treatment effects. This paper we apply uplift modeling to analyze the effect of company adjustment on their financial status, and we treat these adjustment as treatments or interventions in this study. Although there have been extensive studies and application regarding binary treatments, multiple treatments, and continuous treatments, company adjustment are often more complex than these scenarios, as they constitute a series of multiple time-dependent actions. The effect estimation of company adjustment needs to take into account not only individual treatment traits but also the temporal order of this series of treatments. This study collects a real-world data set about company financial statements and reported behavior in Luxembourg for the experiments. First, we use two meta-learners and three other well-known uplift models to analyze different company adjustment by simplifying the adjustment as binary treatments. Furthermore, we propose a new uplift modeling framework (MTDnet) to address the time-dependent nature of these adjustment, and the experimental result shows the necessity of considering the timing of these adjustment.
\keywords{Uplift modeling  \and Deep causal learning \and Financial distress \and Company adjustment}
\end{abstract}

\section{Introduction}

Company adjustments could be the act of changing a manager, change of registered address, change of auditor, etc. Several studies have indicated that there is a correlation between such acts and financial distress and that it is possible to train a bankruptcy prediction model where adjustment acts are treated as normal features~\cite{loderer1989corporate,gillani2018review,yousaf2024corporate}. However, it is often difficult to know which adjustment act leads to a particular financial effect. In this study, our objective is to uncover the causal relationship between adjustment acts and their financial effects, and to estimate the effect of what an adjustment act may lead to.

A precise estimation of the individual treatment effect (ITE) is important in many areas. For example, doctors can provide the most effective treatment to each individual according to the patient's genome, lifestyle, and medical history~\cite{kent2018personalized,jaskowski2012uplift}, and educators can know the most effective way to prevent students from dropping out according to their learning styles, cognitive levels, and interests~\cite{olaya2020uplift}. ITE can also help a company to get to know its customers and to carry out targeted advertising~\cite{zhao2019uplift,liu2023explicit}. We will use adjustment acts as treatments in this study.

Rubin~\cite{rubin1986statistics} proposes the potential Outcome model as a framework to estimate causal effects by controlling for all variables except the treatment. Despite its fundamental importance, meeting the assumptions of this framework, including stable unit treatment value, ignorability, and positivity, presents considerable challenges. The traditional method for estimating individual treatment effects is random controlled trials (RCT). To implement the RCT, the random allocation strategy has to be followed to ensure that there are no confounders during the experiment, so the difference in outcomes can only be caused by treatments. However, the application of RCT is limited because it is time-consuming, costly, and may involve ethical problems. Moreover, RCTs typically focus on the average treatment effect rather than the specific individual treatment effect. While uplift modeling, which refers to a set of machine learning-based techniques used to estimate the effect of a treatment or intervene on an individual~\cite{gutierrez2017causal}, provides a practical way for this problem by estimating the effect of individual treatment from observational data\cite{zhang2021unified,moraes2023uplift}.

Although current application areas of uplift modeling include the study of binary treatments~\cite{kunzel2019metalearners,jaskowski2012uplift,rzepakowski2010decision}, multiple treatments~\cite{rzepakowski2012decision,zhao2017uplift,saito2020cost}, and various other scenarios~\cite{kim2023modeling,parbhoo2021ncore}, the complexity escalates when dealing with multiple time-dependent treatments. Unlike static treatments that produce instantaneous outcomes, dynamic interventions unfold over different time trajectories, presenting a substantial obstacle to previous modeling approaches. 

In this study, the dynamics of company adjustment poses the challenge of estimating its effect on bankruptcy prevention. When faced with financial distress, companies often resort to strategic interventions, ranging from cost-cutting measures to restructuring moves, to avoid bankruptcy. 
In this context, we estimate the effect of the company adjustment on each company and provide some insights for corporate governance to prevent financial distress. To investigate the necessity of considering the timing, we also propose a novel uplift modeling framework for multiple time-dependent treatments. Our framework integrates long short-term memory (LSTM) networks and attention mechanisms to capture the temporal dynamics of company adjustment and its impact on insolvency vulnerability. We empirically validated the effectiveness of the framework and provided insights into its applicability and effectiveness in real-world scenarios. 


Through this effort, we aim to shed light on the dynamic interaction between company adjustment acts and vulnerability to insolvency, ultimately forming strategies for proactive risk management and resilience enhancement. In short, our contributions in this work are:
\begin{itemize}
    \item we propose an efficient framework to estimate the individual effect with multiple time-dependent treatments,
    \item we estimate and analyze the effect of the company adjustment on its financial status,
    \item to the best of our knowledge, this is the first study to apply the uplift modelling in this area.
\end{itemize}

\section{Related work}
Uplift modeling is dedicated to estimate the causal impact of a treatment at an individual level. It has found widespread application in domains such as marketing, healthcare, and personalized recommendations. Rubin's seminal work introduced the potential outcome framework, which forms the basis for causal inference by controlling for all variables except the treatment~\cite{rubin1986statistics}. However, this framework relies on stringent assumptions like stable unit treatment value, ignorability, and positivity that can be challenging to satisfy in real-world scenarios.

Recent research~\cite{forastiere2021identification} has proposed new estimations defining treatment and interference effects using observational data. This sets the groundwork for the use of industrial observational data to estimate uplift. Subsequent studies have increasingly applied deep learning models in causal inference following the publication of Shalit et al.'s study~\cite{shalit2017estimating}. Kunzel et al. introduced meta-learners for uplift modeling, dividing the learning process into two stages: one for predicting outcomes under treatment and another for predicting outcomes under control~\cite{kunzel2019metalearners}. While effective for binary treatments, this approach encounters challenges with more complex treatment scenarios. Numerous studies have been conducted on various types of treatments in subsequent years; we summarize these studies and their target treatments in Table ~\ref{treat}. Despite research on five different types of treatments, multiple time-dependent treatments remain unexplored. Liu et al.\cite{liu2023explicit} addresses the challenge of fully exploiting the interaction between treatment and context information. We designed our multiple time-dependent treatment module based in part on this study. 

\begin{table}[t]
    \centering
    \caption{Description of various treatments and studies. $T$ refers to treatment. $\mathbb{N}$ denotes the set of nonnegative integers. $\mathbb{R}$ denotes the nonnegative set. $k$ refers to the $k_{th}$ treatment. }
    \begin{tabular}{lll}
    \toprule
Types of treatments & Mathematic definition & Studies\\
\midrule
    Binary & $ T \in \{0,1\}$ & \cite{li2023new,kunzel2019metalearners,jaskowski2012uplift,rzepakowski2010decision}\\
    Multiple & $ T \in \mathbb{N}$ & \cite{olaya2020survey,zhao2017uplift,lopez2017estimation}\\
    Continuous &$ T \in \mathbb{R}$ & \cite{kallus2018policy,callaway2024difference,brown2021propensity,kennedy2017non}\\
    Multi-cause  & $(T_1, T_2, ...,T_k), k \in \mathbb{N}$ & \cite{qian2021estimating,ma2021multi,d2019multi,kong2019multi}\\
    Time-dependent & $ (T \mid time) $  & \cite{thomas2020matching,bodnar2004marginal,barile2024causal}\\
    \bottomrule
    \end{tabular}
    \label{treat}
\end{table}

In related academic studies~\cite{altman2008value,liang2016financial,lin2010role}, it has been fully verified that there is a significant correlation between company adjustment acts and the company's financial status. Taştan and Davoudi~\cite{tacstan2015empirical} suggests there are significant relationships between financial distress and corporate governance practices within variables of the number of members on the board of directors of the company, institutional ownership, managerial ownership, CEO duality, and financial leverage. The study~\cite{Daily1994CORPORATEGA} identified the close relationship between governance structure and the possibility of filing for bankruptcy, and highlight the complex interplay between preinsolvency financial, managerial, and governance factors. The study~\cite{darrat2016corporate} finds that larger boards reduce bankruptcy risk in complex firms, while a higher proportion of inside directors lowers bankruptcy risk in firms needing specialist knowledge but increases it in technically unsophisticated firms. Additionally, the influence of corporate governance variables grows stronger as the time to bankruptcy lengthens, indicating that governance changes may be too late to save firms on the brink of bankruptcy. 

However, while the correlation between company adjustments and bankruptcy prediction has been extensively studied, the causal impact of such company adjustment acts on bankruptcy has not yet been fully investigated. When using machine learning or deep learning models to predict bankruptcy, it is common practice to incorporate various relevant factors into the model~\cite{liang2016financial,wang2023augmenting}.  In doing so, these models can make binary classification predictions about whether or not a company will go bankrupt.   Although this approach considers the correlation between multiple features, including company adjustment acts and prediction targets, it focuses solely on identifying patterns without considering causality.   

In practical applications, however, it is not only important to understand the probability of bankruptcy but also essential to gain insight into how to avoid such situations altogether.  Understanding this causal relationship is crucial for developing effective strategies to prevent or mitigate the risk of bankruptcy.   When a company faces credit risk, the people in charge may take some actions such as adjusting people on board or relocating the store to prevent it. If we can know beforehand whether these actions will be effective or not, we can act more confidently. 

The application of uplift modeling to company adjustment analysis is relatively nascent. Most existing studies on uplift modeling focus on simpler, binary treatment scenarios and do not consider the complexities of multiple time-dependent treatments typical in company adjustment. This gap presents an opportunity for the development and application of more sophisticated models that can handle these complexities. 

Recent advancements in machine learning and deep learning provide promising avenues for addressing these challenges. Long short-term memory (LSTM) networks and attention mechanisms, for example, have shown great potential in capturing temporal dependencies and interactions in sequential data.  These techniques can be leveraged to develop uplift models that account for the dynamic nature of company adjustment and their impact on financial outcomes. Our proposed model, MTDnet, aims to fill this gap by integrating these advanced techniques to handle multiple time-dependent treatments in company adjustment analysis.  By doing so, we seek to provide more accurate and actionable insights into how different company actions influence financial stability and the likelihood of bankruptcy.

\section{Methodology}

\subsection{Dataset}

The dataset utilized for the experiments is a publicly available dataset obtained from the Luxembourg Business Register\footnote{\url{www.lbr.com}}. Companies are mandated to disclose files pertaining to their registered information, financial reports, managerial changes, and other essential details in Luxembourg. The data collected spans from 2011 to 2021, with a subsequent focus on small and medium companies within our research scope. Due to favorable tax policies, there is a notable presence of financial companies and large corporate headquarters in Luxembourg; however, this may not be representative of the general market conditions in other countries, so we exclude the financial-related companies.

\begin{figure}[t]
\includegraphics[width=\textwidth]{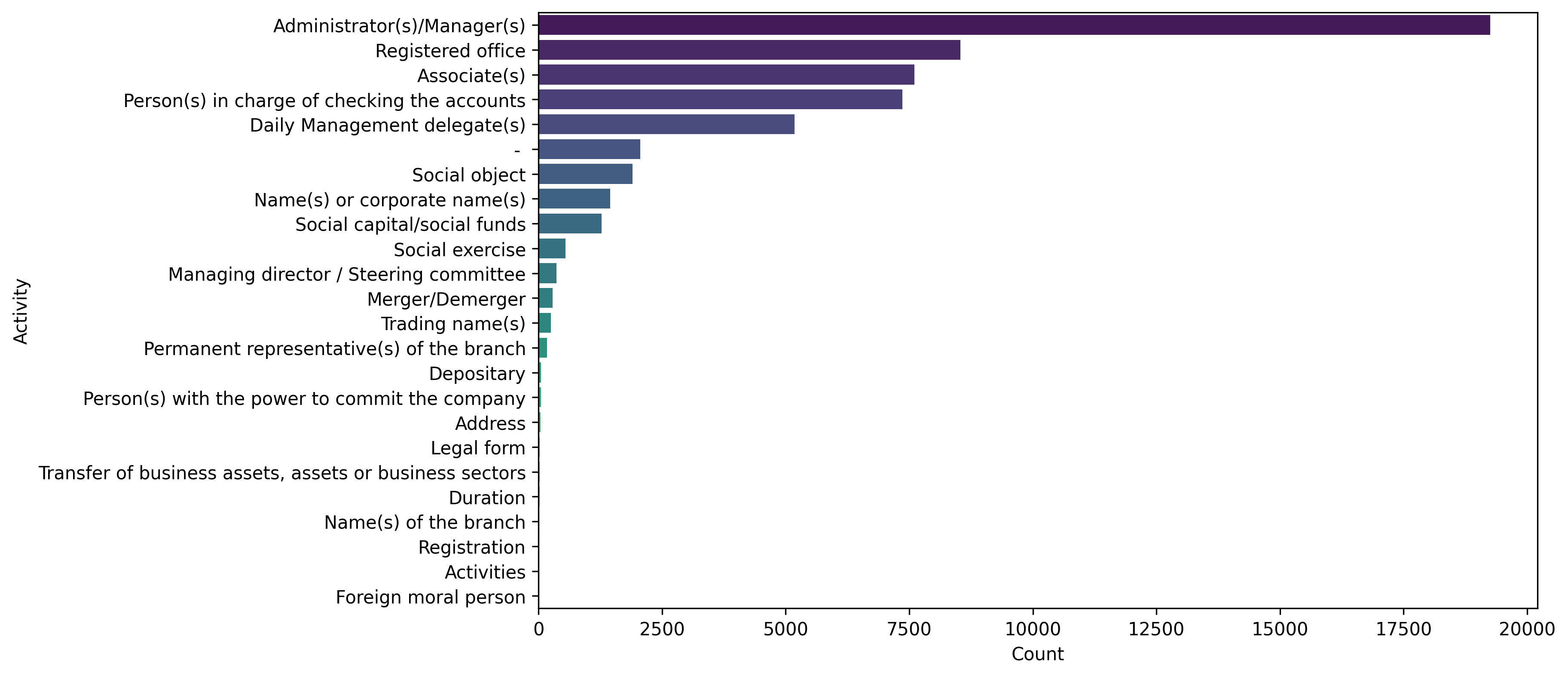}
\caption{Distribution of company adjustment acts} \label{activity}
\end{figure}

The dataset comprises basic information, financial statements, and company adjustment acts  derived from company-reported files. For analysis and uplift modeling purposes, we have selected the most recent financial statements of each company along with their adjustment acts  occurring between the release date of these statements and their current financial status (active or bankrupt). Basic information encompasses legal form, operational duration, and sector classification. Figure~\ref{activity} illustrates the distribution of different adjustment acts  observed in this study. The Luxembourg Business Register classifies company adjustments into 24 distinct categories; however, it should be noted that these acts exhibit significant imbalance in distribution. The most frequent acts are related to the changing of managerial people. 

\subsection{Problem formulation and data mapping}
To investigate the effects of company adjustment acts on financial stability, we examine the company's likelihood of bankruptcy as the dependent variable and company adjustment acts as treatments. The independent variables include a company's financial statements and basic information.

We apply the uplift model to estimate the ITE of each company under specific treatments. By conducting the counterfactual predictions $ITE = (y_i|(T=k_t) - y_i|(t=0))$, we can determine whether a treatment has an effect on the outcome. In our context, $y_i$ refers to company$i$ , $T=k_t$ denotes that a company implements an adjustment act  $k$ at $t$ moment and $T=0$ indicates that a company does not take any actions. A positive ITE suggests that adjustment acts have a positive effect on preventing bankruptcy, while a negative value indicates otherwise.

The research questions (RQ) defined in this study are:

RQ1: What types of company adjustment acts help improve financial status and prevent bankruptcy?

RQ2: Should the sequence and timing of company adjustment acts be considered when measuring their impact on financial stability?

For RQ1, in order to study effective types of adjustment acts, we transform original treatments into binary treatments. Due to limited samples for some company adjustment, it is challenging for the uplift model to estimate each type individually. Therefore, we restructure and categorize these adjustment into four datasets for uplift models:
\begin{itemize}
    \item Basic binary treatment: where $T=0$ if no reported adjustment exists; otherwise $T=1$
    \item Personnel-related treatment: where personnel-related acts are classified as $T=1$; otherwise $T=0$
    \item Information-related treatment: where changing information-related or business-related acts  are categorized as $T=1$; otherwise $T=0$
    \item Other activities: all other unclassified adjustment acts are categorized as $T=1$; otherwise $T=0$
\end{itemize} 

From Table~\ref{dataset}, we observe the distribution across these four datasets with $y=1$ indicating companies going bankrupt within one year and $y=0$ representing companies still operating. We implement the uplift models on these datasets and compare the uplift of these four different types of treatments.

For RQ2, we propose an uplift model to address the problem arising from the special data structure of company adjustment acts. We compare its performance with traditional uplift models used for binary or multiple treatments in order to assess whether considering different treatment sequences is necessary.

We randomly sampled the training and test sets in the ratio of 7:3, and the metrics values in all the experimental results are the values of the test set.

\begin{table}[htbp!]
    \centering
    \caption{Description of the four datasets for uplift models}
    \label{dataset}
\begin{tabular}{c|c|c|c|c|c|c|c|c}
\toprule
    & \multicolumn{2}{|c|}{Basic binary} & \multicolumn{2}{|c|}{Personnel} & \multicolumn{2}{|c|}{Information} & \multicolumn{2}{|c}{Other} \\
    \midrule
    & T=0   & T=1   & T=0   & T=1   & T=0   & T=1   & T=0   & T=1   \\
    \midrule
y=0 & 21820 & 17974 & 25600 & 14194 & 34417 & 5377  & 25600 & 14194 \\
y=1 & 2281  & 4529  & 3589  & 3221  & 5347  & 1463  & 3589  & 3221  \\
\bottomrule
\end{tabular}
\end{table}

\subsection{MTDnet Framework}
\begin{figure}
    \centering
    \includegraphics[width=\linewidth]{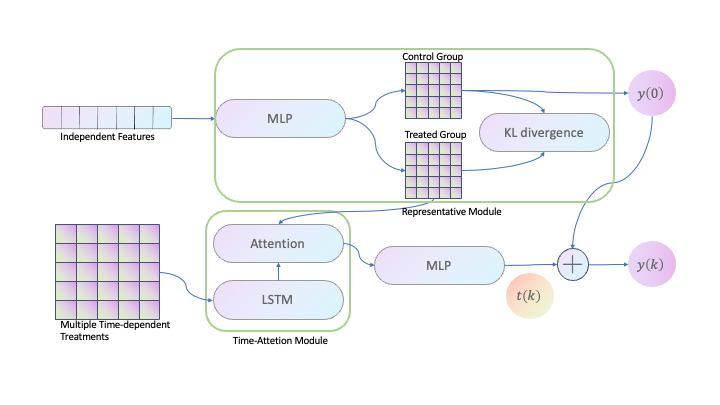}
    \caption{Framework of multiple time-dependent uplift modelling}
    \label{frame}
\end{figure}

The diagram in Fig.~\ref{frame} illustrates the comprehensive framework to estimate the uplift with multiple time-dependent treatments. The input data shown in the figure represent contextual features, which are one-dimensional array-like data. The matrix-like data of multiple time-dependent treatments describe the time steps and multiple treatments. This framework consists of two main components: the representative module and the time-attention module. The uplift model is a two-head network model; one head estimates the label value when there is no treatment, while the other head estimates the label value when treatments are implemented. We use $L_C$ to denote the loss of control group ($t=0$, where $t$ denotes treatment) and $L_T$ to denote the loss of treated group ($t=k$). In both cases, a lower loss indicates a more precise estimate. $L_D$ measures the distance between the control group and treated group using Kullback-Leibler (KLD) divergence. A smaller KLD signifies higher similarity between control and treated groups, indicating that these experiments resemble randomized controlled trials (RCTs) more closely.
\begin{eqnarray}\label{eq:ld}
L_C = loss\{y(t=0), \hat{y}(t=0)\}
\end{eqnarray}
\begin{eqnarray}\label{eq:lt}
L_T = loss\{y(t=k), \hat{y}(t=k)\}
\end{eqnarray}
\begin{eqnarray}\label{eq:kld}
L_D = KLD\{dist.|control, dist.|treated\}
\end{eqnarray}

In general, the loss function of this model is 
\begin{eqnarray}\label{eq:total}
total \ loss = L_C + L_T + L_D
\end{eqnarray}

\paragraph{Representative Module}
For all the independent features, we put them in the representative module. We used multilayer perceptrons to capture the characteristics of both the control group and the treatment group. Independence is an important assumption in causal inference, which means that the potential causal variable should be independent of other potential causes of the outcome. This means that there should be no other factors that simultaneously affect both the potential cause and the outcome.
When handling the observational data, we try to mimic the randomized controlled trials in order to have fewer selection biases. After sharing the same layers, we use the KLD to calculate the distance of the distribution of control group and treatment group. The lower KLD means that the distribution of the control group and the treatment group is more like, the selection bias will be less.

\paragraph{Multiple Time-dependent Treatment module}
In this module, we adopt the long and short-term networks (LSTM) to capture time-related characteristics of multiple time-dependent treatments. LSTM is flexible to handle time series data. After the LSTM layer, we apply the attention theory to optimize the treatment representation by considering the background features. We calculate the different attention weights of the treatments and multiply the weights with the treatments to get the interactive results.

\subsection{Metrics}
We adopt four commonly used metrics to evaluate the performance of the proposed model~\cite{diemert2018large,devriendt2020learning}: Average Uplift, Qini Score, AUUC, and Uplift at 30\%.

\paragraph{Average Uplift} refers to the average effect of a treatment or intervention across a population. It measures the difference in outcomes between those who received the treatment and those who did not, averaged across all individuals.

\paragraph{Qini Score} computes normalized Area Under the Qini coefficient curve from prediction scores. By computing the area under the Qini curve, the curve information is summarized in one number. For binary outcomes, the ratio of the actual uplift gains curve above 
the diagonal to that of the optimum Qini curve.

\paragraph{AUUC} computes normalized Area Under the uplift curve from prediction scores. There are many different ways to calculate the AUUC value, in this paper, we adopt the formulation from~\cite{gutierrez2017causal}. By computing the area under the uplift curve, the curve information is summarized in one number. For binary outcomes the ratio of the actual uplift gains curve above the diagonal to that of the optimum uplift curve.

\paragraph{Uplift at 30\%} computes uplift at first 30\% observations by uplift of the total sample. After ordering the data by uplift prediction, the difference of conversions between control group and treatment group can be get.

\subsection{Uplift models}
\label{uplift-models}
We have chosen two models from meta-learner and three other models according to the study by Gutierrez and Gérardy~\cite{gutierrez2017causal}.

\paragraph{S-Learner (Single Model Learner)} is a fundamental model in uplift modeling that estimates the individual treatment effect by employing a single model. It is represented by the following equation:
\[\hat{Y}_i = \hat{Y}(X_i,1)-\hat{Y}(X_i,0)\]
Here, \(\hat{Y}_i\)denotes the uplift score for individual \(i\), \(\hat{Y}(X_i,1)\) represents the predicted outcome for individual \(i\) under treatment, and \(\hat{Y}(X_i,0)\) represents the predicted outcome for individual \(i\) without treatment.

\paragraph{T-Learner (Two Model Learner)} employs two separate models to estimate the expected outcomes for the treatment and control groups, respectively, and calculates the difference between them. It is expressed by the following equation:
\[\hat{Y}_i = \hat{Y}_1(X_i)-\hat{Y}_0(X_i)\]
Here, \(\hat{Y}_i\) represents the uplift score for individual \(i\), \(\hat{Y}_1(X_i)\) denotes the predicted outcome for individual \(i\) under treatment, and \(\hat{Y}_0(X_i)\) denotes the predicted outcome for individual \(i\)without treatment.

\paragraph{CEVAE (Causal Effect Variational Autoencoder)} is a probabilistic graphical model-based uplift model that estimates individual uplift scores through a variational autoencoder (VAE). It involves complex equations related to the encoder and decoder structures for learning individual latent representations and causal relationships in the context of treatment effects. In essence, CEVAE aims to disentangle the latent factors that influence both treatment assignment and outcome, allowing for more accurate estimation of causal effects. 

\paragraph{DragonNet} consists of two neural networks: one to predict the outcome of the treatment group and the other to predict the outcome of the control group. The two networks share part of the structure (called the "shared representation"), and then each has its own output layer. By comparing the output of the two networks, Dragonnet can estimate individual processing effects.
A key feature of Dragonnet is that it uses a special loss function designed to directly optimize the quality of the estimated individual processing effects. This allows Dragonnet to provide more accurate estimates than traditional causal inference methods when dealing with complex, high-dimensional data.

\paragraph{Ganite}is a generative adversarial network (GAN)-based method for estimating ITE. It uses two neural networks: a generator and a discriminator. The generator's task is to generate possible therapeutic effects, while the discriminator's task is to distinguish between generated therapeutic effects and real therapeutic effects. In this way, the generator is trained to generate therapeutic effects that are closer to the real thing. This is achieved by having the generator generate a distribution of the therapeutic effects, rather than just a point estimate. This allows GANITE to provide richer information to help decision-makers understand the uncertainty of treatment effects.

\subsection{Implementation Details}
We use the PyTorch (version 2.2.1) to build the network. For uplift models Dragonnet and CEVAE, we use the CausalML package~\cite{chen2020causalml} with the default settings. For S-learner and T-learner, we use the lightgbm model as the estimator for the model with the default settings. For Ganite, we adopt the settings from the paper~\cite{yoon2018ganite}.
For tuning our proposed model, we use the GridSearch method and the early stop condition that patience of 8. The hyperparameters can be seen from Table~\ref{params}.

\begin{table}[]
    \centering
    \caption{Scope of hyperparameters}
    \begin{tabular}{ll}
    \hline
      Hyperparameter  & Range \\
      \hline
        Batch size & {32,64,128}\\
        Number of Epoch & {50,100,150} \\
        Learning rate & {0.0001, 0.00001,0.000005}\\
        L2 & {$1e^-4$,$1e^-5$,$1e^-6$}\\
        Hidden size & {$2^5$,$2^6$,$2^7$,$2^8$}\\
        Output size & {$2^3$,$2^4$,$2^5$}\\
        \hline
    \end{tabular}
    \label{params}
\end{table}

\section{Results and Discussion}
In this section, the results and findings of the previous defined research questions are discussed.

\subsection{RQ1: Identifying Effective company adjustment}
\begin{figure}[htp]
\centering
\includegraphics[width=\linewidth]{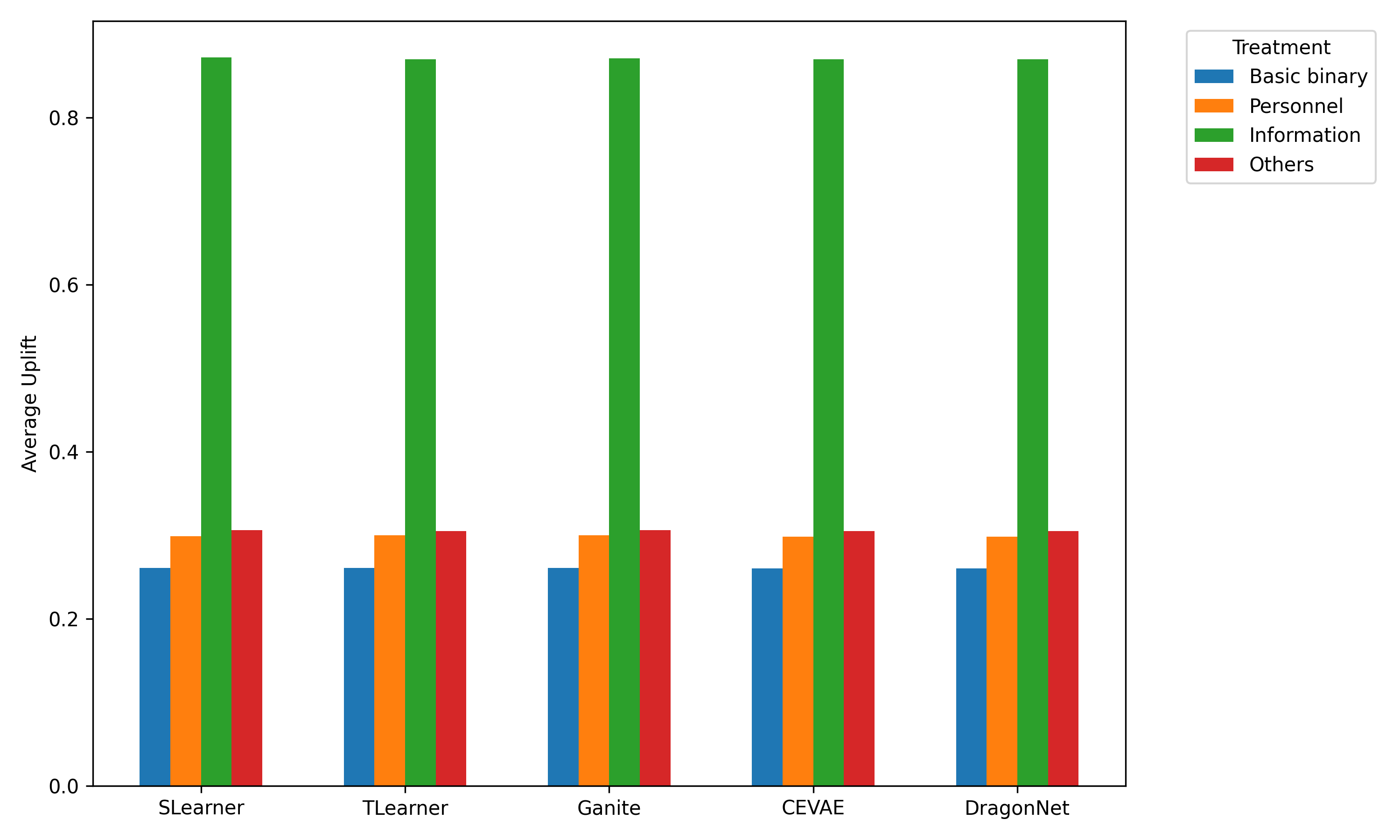}
\caption{Uplift comparison for different adjustment}
\label{fig:uplift}
\end{figure}

We evaluate the uplift for different categories of treatments by the uplift models mentioned in section~\ref{uplift-models}. Fig~\ref{fig:uplift} indicates the overall results for different uplift models show the same conclusion. The dataset of information and business-related treatment has the absolute advantage for the uplift over the other three types of treatment according to the results of all the five uplift models. The basic binary treatments have the least uplift among the four datasets, and this may because the treatments have the conflict with each other, that the mixed treatments perform the worst.

Personnel-related treatments contain a relatively large number of adjustment acts  that  do not gain a lot uplift for all the models. We think the reason may be it is challenging to standardize the treatment as it is not possible to have the same or similar human being. Even if it is the same person, we cannot say (s)he will make the same decision under the same scenario. Therefore, this kind of adjustment cannot have a good performance in uplift modelling.

\subsection{RQ2: Necessity of Considering the Sequence of Treatments}
We implement the experiments on the LBR data and compare our proposed model with six other representative uplift models mentioned in previous section. We have not been ablt to find any open source datasets which also involve the multiple time-dependent treatments. Therefore, we cannot evaluate  our model on other datasets for comparison. Also, there are no published uplift models that designed for handling  multiple time-dependent treatments. 

The original treatment contains three dimensions: value of treatment, type of treatment and timing of treatment. We reshape the structure of treatment in order to evaluate if it is necessary to consider all the three dimensions. For the binary treatment scenario, we ignore the type and timing of the treatment. Therefore, if $T=1$ means this company takes some actions. For multiple treatment scenario, we ignore the timing and the company can have multiple adjustment without considering the sequence of adjustment. 

Although MTDnet has a lower value for uplift at 30\% which indicate that this is not as effective as others in the top 30\% of the target group, MTDnet outperforms the other treatment scenarios for the metrics AUUC and Qini which suggests it's better identifies treatment effects across the entire dataset and is able to accurately predict which individuals were positively affected by the treatment over a wider range(see Table~\ref{result}). In the future, we could consider combining the strengths of other models to build hybrid models to achieve better results both globally and locally.

Moreover, we find the network-based models (CEVAE, DragonNet and Ganite) perform much worse than its published results, while machine learning-based models are more robust. We think this is because network-based models normally need more delicately tuned hyperparameters. As the authors~\cite{olaya2020survey} pointed out, there is no model that always perform the best for all the context and problems. 

\begin{table}[h]
    \centering
    \caption{Results of the experiments for evaluating the necessary of considering treatment sequence. The suffix of the model name represents the model is trained with different treatments. "-bi" refers to binary treatments, "-multi" refers to multiple treatments, and "-original" refers to multiple time-dependent treatments.}
    \begin{tabular}{llll}
    \hline
       Model  & Uplift at 30\% & AUUC & Qini \\
       \hline
        S-learner-bi	&	0.0161	&	0.0272	&	0.0478	\\
        T-learner-bi	&	\textbf{0.0489}	&	0.0422	&	0.0709	\\
        CEVAE-bi	&	0.0049	&	0.0094	&	0.0169	\\
        DragonNet-bi	&	0.0000	&	0.0008	&	0.0013	\\
        Ganite-bi	&	-0.0037	&	0.0206	&	0.0364	\\
        \hline							
        Ganite-multi	&	0.0022	&	-0.0160	&	-0.0289	\\
        \hline							
        MTDnet-original	&	0.0067	&	\textbf{0.0589}	&	\textbf{0.1880}\\
        \hline
    \end{tabular}
    \label{result}
\end{table}



\section{Conclusion}

In this study, we have explored the application of uplift modeling to analyze the impact of company adjustment on financial stability, specifically focusing on the prevention of bankruptcy. Our research addressed two main questions: identifying effective company adjustment and investigating the necessity of considering the sequence of treatments.

We evaluated the uplift for different categories of treatments using several uplift models. The results, depicted in Fig~\ref{fig:uplift}, consistently show that information and business-related treatments had a significant advantage over other types of treatments across all models. This suggests that interventions focused on improving information flow and business operations are the most effective in enhancing company financial stability.
On the other hand, basic binary treatments showed the least uplift, likely due to conflicting actions within this category, resulting in mixed and suboptimal outcomes. Personnel-related treatments also demonstrated lower uplift, possibly because of the inherent variability and unpredictability associated with human behaviors, making standardization and prediction challenging.

To address the second research question, we compared our proposed model, MTDnet, with six other uplift models using data involving multiple time-dependent treatments. Given the lack of open-source datasets with similar characteristics, our experiments were confined to our dataset. The results showed that MTDnet outperformed other models in terms of AUUC and Qini metrics, indicating the necessity of considering the value, type, and timing of treatments for accurate uplift estimation.

Our findings also revealed that network-based models performed worse than their published results, whereas machine learning-based models demonstrated more robustness. This discrepancy might be attributed to the need for more precise hyperparameter tuning in network-based models, as highlighted by previous research.

Through this study, we contribute to the field of uplift modeling by proposing an efficient framework to estimate individual effects with multiple time-dependent treatments, demonstrating the superiority of MTDnet in handling complex, sequential company adjustment, highlighting the importance of considering all dimensions of treatments (value, type, and timing) for accurate effect estimation.
Our research underscores the dynamic interplay between company adjustment and financial stability, offering valuable insights for proactive risk management and resilience enhancement. By identifying the most impactful interventions and validating the necessity of sequence consideration, we provide a foundation for developing more effective strategies to prevent bankruptcy.

In summary, this study not only advances the application of uplift modeling in company finance but also emphasizes the critical role of temporal dynamics in understanding and improving company resilience. Future research can build on these findings to explore broader datasets and refine the proposed models for even greater accuracy and applicability.




\bibliographystyle{splncs04}
\bibliography{reference}

\end{document}